\documentclass[12pt,preprint]{emulateapj}

\def\msun{{\rm M_{\odot}}}

\def\chisq{{\chi^{2}}}

\def\H0{{\rm ~km~s^{-1}~Mpc^{-1}}}

\def\msun{M_{\rm \odot}}

\def\deg{^\circ}

\shorttitle{GRB 050911: a black hole - neutron star merger or a naked GRB}
\shortauthors{K.L. Page et al.}

\begin{document}


\title{GRB 050911: a black hole - neutron star merger or a naked GRB}


\author{K.L. Page\altaffilmark{1}, A.R. King\altaffilmark{1}, A.J. Levan\altaffilmark{2}, 
  P.T. O'Brien\altaffilmark{1}, J.P
  Osborne\altaffilmark{1}, S.D. Barthelmy\altaffilmark{5}, A.P. Beardmore\altaffilmark{1}, 
  D.N. Burrows\altaffilmark{3}, S. Campana\altaffilmark{4},
  N. Gehrels\altaffilmark{5}, J. Graham\altaffilmark{6}, M.R. Goad\altaffilmark{1},
  O. Godet\altaffilmark{1}, Y. Kaneko\altaffilmark{7}, J.A. Kennea\altaffilmark{3},
  C.B. Markwardt\altaffilmark{5}, D.E. Reichart\altaffilmark{8},
  T. Sakamoto\altaffilmark{5} \&  N.R. Tanvir\altaffilmark{2}}

\altaffiltext{1}{Department of Physics and Astronomy, University of Leicester, Leicester, LE1 7RH, UK}
\altaffiltext{2}{Centre for Astrophysics Research, University of
  Hertfordshire, Hatfield, AL10 9AB, UK}
\altaffiltext{3}{Department of Astronomy \& Astrophysics, 525 Davey Lab, Pennsylvania State University, University Park, PA 16802, USA} 
\altaffiltext{4}{INAF, Osservatorio Astronomico di Brera, Via E. Bianchi 46,
I-23807, Merate (LC), Italy}
\altaffiltext{5}{NASA/Goddard Space Flight Center, Greenbelt, MD 20771, USA}
\altaffiltext{6}{Space Telescope Science Institute, 3700 San Martin Drive, Baltimore, MD21218, USA} 
\altaffiltext{7}{University of Alabama, National Space Science and Technology
  Center, Huntsville, AL 35805, USA}
\altaffiltext{8}{Department of Physics and Astronomy, University of North
  Carolina at Chapel Hill, Chapel Hill, NC 27599, USA} 

\email{kpa@star.le.ac.uk}

\begin{abstract}

GRB~050911, discovered by the {\it Swift} Burst Alert Telescope, was not seen
4.6~hr later by the {\it Swift} X-ray Telescope, making it one of the very
few X-ray non-detections of a Gamma-Ray Burst (GRB) afterglow at early times. The $\gamma$-ray light-curve shows at least three peaks, the
first two of which ($\sim$T$_0-0.8$ and T$_0+0.2~\rm s$, where T$_0$ is the trigger time) were
short, each lasting 0.5~s. This was followed by later emission
10--20~s post-burst. The upper limit on the unabsorbed
X-ray
flux was 1.7~$\times 10^{-14}~\rm erg~cm^{-2}~s^{-1}$
(integrating 46~ks of data taken between 11 and 18 September),
indicating that the decay must have been rapid. 
All but one of the long
bursts detected by {\it Swift} were above this limit at
$\sim$4.6~hr, whereas the afterglows of short
bursts became undetectable more rapidly. 
Deep observations
with Gemini also revealed no optical afterglow 12~hr after the burst, down to $r=24.0$
(5$\sigma$ limit). We speculate that GRB~050911 may have been formed through a compact object (black
hole-neutron star)
merger, with the later outbursts due to a longer disc lifetime linked to a large
mass ratio between the merging objects. Alternatively, the burst may have occured in a low
density environment, leading to a weak,
or non-existent, forward shock -- the so-called `naked GRB' model.

\end{abstract}


\section{Introduction}

The bimodality in Gamma-Ray Burst (GRB) durations has long been recognised (e.g. Kouveliotou
et al. 1993) with the 90\% $\gamma$-ray emission interval (T$_{90}$) peaking
around 0.3 and 30~s, with a minimum at two seconds.
The short duration
bursts also typically
exhibit systematically harder emission than
the longer ones (Kouveliotou et al. 1993). The revolution which 
transformed the study of long duration bursts via the identification
of afterglows and host galaxies at cosmological redshifts has only
just reached the short bursts. The recent
discoveries of short burst afterglows in several cases (e.g. GRB~050509B:
Gehrels et al. 2005, Bloom et al. 2005; GRB~050709: Covino et
al. 2005, Fox et al. 2005b, Hjorth et al. 2005a, Villasenor et al. 2005; GRB~050724: Barthelmy et al. 2005b;
GRB~050813: Fox et al. 2005c) and the association of these with host galaxies of various
morphological types (including ellipticals) indicate that short GRBs have a
different origin from the longer duration bursts. They are typically found at 
lower redshifts (e.g. Bloom et al. 2005; Berger et al. 2005; Tanvir et al.
2005) and have isotropic energies three orders of magnitude below those of the
long GRBs. The definitive lack of detection of a supernova related to
GRB~050509B (Hjorth et al. 2005b) supports the difference between long and short
bursts. The
observations to date are in line with what may be expected from GRBs
occuring via compact object mergers [neutron star-neutron star (NS-NS) or
 black hole-neutron star (BH-NS)]. 

The two burst populations clearly overlap in the hardness duration
parameter space and it is interesting to ask what distinguishes
the different classes for the cases where classification simply via $\rm T_{90}$
is ambiguous. At least one of the
short GRBs found by {\it Swift} -- GRB~050724; Barthelmy
et al. 2005b -- has softer emission beyond the expected T$_{90}$ of 2~s. In fact, these data showed that 
the distinction between long and short bursts is partly instrument-dependent.
 Using {\it Swift} data, the $\gamma$-ray
light-curve of that burst was found to consist of an initial hard spike
(lasting about 0.25~s), followed by another peak at T$_0+1.1~\rm s$; this section of the light-curve
would have classified GRB~050724 as a short burst. However, faint, softer
emission was detected by the  {\it Swift} Burst
Alert Telescope (BAT) out to 140~s after the
trigger. Simulations showed that BATSE (the Burst And Transient Source
Experiment) would not have detected this softer pulse at more than
0.3$\sigma$, obtaining T$_{90} \sim 0.43~\rm s$. Similar behaviour was seen for GRB~050709 (Fox et al. 2005b) and both
Lazzati, Ramirez-Ruiz \& Ghisellini (2001) and Connaughton
(2002) investigated such `tails' in BATSE short burst data. Norris \& Bonnell (2005) have looked at short bursts with extended emission, finding
that they can be differentiated from long bursts by having spectral lags
consistent with zero for their initial spike emission.

Huang et al. (2005) discuss GRB~040924. The duration of this burst (T$_{90}
\sim 1.2~\rm s$)
places it in the short category, but the authors conclude that it might belong at the short end of the long
GRB distribution. The presence of a supernova signature in this burst
strengthens their assertion (Soderberg et al. 2005). Likewise, GRB~000301C
(Jensen et al. 2001) had $\rm T_{90}~=~\rm 2.0~s$, but all the properties of a
long duration burst (e.g., starforming host galaxy, optical and radio
detections of the afterglow and high redshift). These observations provide
evidence that the long-duration population extends to at least 1-2~s.

GRB~050911 also appears to be a candidate for a burst whose classification via
$\rm T_{90}$ is unclear. Although $\rm T_{90}$ places it in the
long burst category, any later X-ray or optical emission decayed rapidly
and/or was extremely faint, characteristics more common for short
bursts (see, e.g., Gehrels et al 2005, Fox et al. 2005c).  
This Letter presents the {\it Swift} and ground-based observations of
the burst (Sections~\ref{swift} and \ref{followup}), finding that it
was only detected in $\gamma$-rays. Explanations for the lack of X-ray
emission are explored in Section~\ref{disc}.

\section{{\it Swift} observations}
\label{swift}

The {\it Swift} Gamma-Ray Burst Explorer (Gehrels et al. 2004) is a
multi-wavelength mission designed to detect and study GRBs.
The
observatory consists of three instruments -- the wide-field $\gamma$-ray BAT (Barthelmy et al. 2005a) and the two narrow-field
instruments: the X-ray and Ultraviolet/Optical telescopes (XRT; Burrows et
al. 2005 and UVOT; Roming et al. 2005). 


GRB~050911 was located by the BAT (trigger number 154630) on 11 September 2005 at 15:59:34 UT (Page et al. 2005a). The
ground-calculated BAT position (Tueller et al. 2005) was RA = $00^{h} 54^{m}
52.4^{s}$, Dec  = $-38\deg 51\arcmin 42.8\arcsec$ (J2000), with an uncertainty of
2.8$\arcmin$ (radius, 90\% containment). $\rm T_{90}$ was determined to be
$\sim 16~\rm s$, while $\rm T_{50} \sim 13.7~\rm s$; 
the T$_{90}$ fluence over 15--150~keV is $\sim 3 \times 10^{-7}~\rm
erg~cm^{-2}$, placing it at the lower end of the BAT distribution.

Due to a combination of the Earth-limb observing constraint and a temporary problem with
the star trackers, the first XRT and UVOT observations of the position of GRB~050911
started 4.6~hr after the burst trigger, when a 6.3~ks
exposure was obtained. Table~\ref{obs} gives the
details of this and the subsequent X-ray observations of the burst. Throughout this Letter, errors are
given at the 90\% significance level unless otherwise stated.

\begin{table}
\begin{center}
\caption{{\it Swift} XRT observing timeline for GRB~050911 \label{obs}}
\begin{tabular}{ccccc}
\tableline
&\\
Obs. ID & Date & Start UT & Stop UT & Exposure (s)\\ 
\\ 
\tableline
00154630000 & 2005-09-11 & 20:33:00 & 01:08:52 & 6343 \\
00154630001 & 2005-09-12 & 14:16:31 & 20:48:30 & 9954\\
00154630002 & 2005-09-14 & 01:31:14 & 22:43:59 & 13043\\
00154631003 & 2005-09-18 & 06:33:52 & 22:51:57 & 17137\\
\tableline
\end{tabular}

\end{center}
\end{table}

\vspace{-10pt}
\subsection{$\gamma$-ray data}

\begin{figure}
\begin{center}
\includegraphics[clip, width=5.0cm,angle=-90]{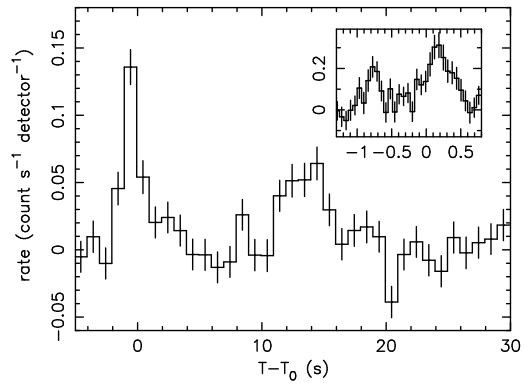}
\caption{The BAT light-curve over 15--350~keV. The main plot shows both the
  initial short burst and the longer, later emission, while
  the inset zooms in on the two short peaks.}
\label{BATlc}
\end{center}
\end{figure}

Fig.~\ref{BATlc} plots the BAT light-curve, showing two short spikes, each about 0.5~s in duration, close to $\rm T_{0}-0.8$ and
$\rm T_{0}+0.2$~s (where T$_{0}$ is the burst trigger time), followed by a slow
rise and fall between $\rm T_{0}+10$ and $\rm T_{0}+20$~s.  

The T$_{90}$ BAT spectrum can be well fitted over 15--150~keV with a simple power-law, of photon
index $\Gamma = 1.90^{+0.33}_{-0.31}$ ($\chisq$/dof~=~54/56). Fitting either
a Band function (Band et al. 1993) or a cut-off power-law does not improve the
fit.

Separate spectra for the time periods covering the first two spikes in the
light-curve ($\rm T_0-4~$ to $\rm T_0+6$~s) and the longer, broader peak
($\rm T_0+6~$ to $\rm T_0+26$~s) were extracted, to determine whether the later emission
appeared softer, as was the case for the two short bursts GRB~050709 and
GRB~050724 (Barthelmy et al. 2005b; Boer et al. 2005). Although there is some
evidence for spectral softening with time ($\Gamma_{\rm early} = 1.82^{+0.73}_{-0.67}$;
$\Gamma_{\rm later} = 2.29^{+1.23}_{-0.49}$), within the errors the photon
indices are consistent. However, as Barthelmy et al. (2005b) found for
GRB~050724, simulations show it is likely that the later, weaker peak would
not have been detected significantly by BATSE ($< 1\sigma$); it should be
noted, though, that the first peak would also have been marginal ($<
3.5\sigma$), so BATSE may not have triggered on the event at all. 

The hardness ratio [S(50-100~keV)/(25-50~keV)] is
consistent with the ranges observed for both short and long bursts detected by
BATSE and {\it Swift}. Following
Norris \& Bonnell (2005), the spectral lag measurement for the initial peak is
consistent with zero, as found for short bursts.

\vspace{-10pt}
\subsection{X-ray data}

Only one source was detected in the BAT error circle by the XRT (Page et al. 2005b), but
this was found to be constant over time and therefore was discounted as the
afterglow of GRB~050911 (Page et al. 2005c).

For any other object in the BAT error circle, the 3$\sigma$ upper limit for
the count-rate is $4.0 \times 10^{-4}~\rm count~s^{-1}$ evaluated over the entire 46~ks exposure (see Table~\ref{obs}).
The corresponding unabsorbed flux limit is
$1.7 \times 10^{-14}~\rm erg~cm^{-2}~s^{-1}$ over 0.3--10~keV, assuming a
Crab-like spectrum with a Galactic absorbing column of $2.7 \times10^{20} \rm cm^{-2}$ (Dickey \& Lockman 1990). Considering just the initial 6.3~ks exposure at 4.6~hr
after the trigger, the
upper limit is $1.25 \times 10^{-3}~\rm count~s^{-1}$ ($5.2\times 10^{-14}~\rm erg~cm^{-2}~s^{-1}$, unabsorbed, over 0.3--10~keV).

\begin{figure}
\begin{center}
\includegraphics[clip, width=5.0cm,angle=-90]{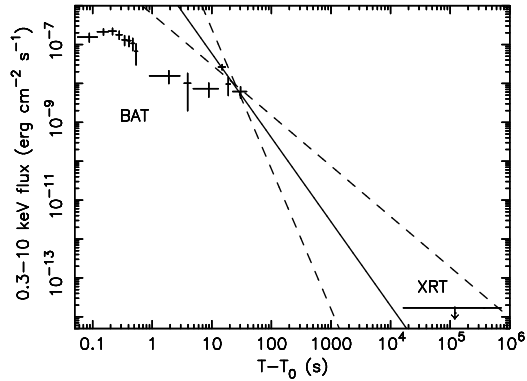}
\caption{The BAT flux light-curve and XRT upper limit, showing an
  extrapolation of the power-law fit to the last peak in the BAT.}
\label{bat-xrt}
\end{center}
\end{figure}

Fig.~\ref{bat-xrt} shows the BAT light-curve extrapolated into the
0.3--10~keV band (assuming a slope of $\Gamma = 1.9$) compared to the XRT upper limit. Fitting a power-law decay to
the later peak in the BAT light-curve gives a slope of
$\alpha = 2.2^{+1.4}_{-0.9}$; the solid and dashed lines show
$\alpha = 2.2$ and the upper and lower limits respectively. The X-ray upper
limit is consistent with the extrapolation of the fit to the end of the BAT
data.

\vspace{-10pt}
\subsection{Optical and UV data}

No new sources were found in the field by the UVOT,
with 3$\sigma$ upper limits on the magnitude of between 20--21 for the optical and UV
filters; the exposures were between 600 and 900~s in duration
(Breeveld et al. 2005).

\section{Ground-based optical observations}
\label{followup}

We observed the BAT position (Tueller et al. 2005)
with the Gemini South Telescope and GMOS instrument over two epochs
(midpoint times of 2005-09-12 04:05:14 UT and 2005-09-15 05:33:18 UT). Images were obtained in the $r$-band, with exposure times of 1500s (5x300s) and 3000s
(10x300s); 5$\arcsec$ dithers were made between subsequent exposures
to cover the chip gaps in the GMOS detectors, and the field of
view of 5.5$\arcmin$ covered the complete refined
BAT error circle. Note that the conditions were sub-optimal, with relatively
poor seeing ($\sim$ 1.2 \arcsec).
No new sources were seen in the GMOS images in comparison to the
Digitized Sky Survey; however, as the Gemini observations are
significantly deeper, a further search for variability was conducted
by performing a point spread function matched image subtraction
of the two epochs using the ISIS-II code of Alard \&
Lupton (1998). To estimate the sensitivity to variable objects, 
a set of artificial stars were created in the first epoch image
based on a zeropoint obtained from comparison to the USNO catalogue,
and the FWHM equal to that of stellar objects within the field, and the
difference image investigated. Using this technique a conservative
5$\sigma$ limiting magnitude of $r$~=~24.0 was determined for any variable object
within the BAT error circle 12~hr after the trigger.

No follow-up detections of GRB~050911 were reported through the GCN (Tristram et
al. 2005; Castro-Tirado et al. 2005; Berger \& Boss 2005).

\section{Discussion}
\label{disc}

The non-detection of an X-ray afterglow is
very unusual for {\it Swift} bursts. Before GRB~050911, only four other GRBs (050416B, 050502A, 050709 and 050509C) had been observed with
the XRT with no afterglow detection; however, these were first observed with
the XRT after 86~hr, 11 days, 39~hr and 9 days respectively, all significantly later than for
GRB~050911. As of 1 November 2005, {\it Swift} slewed to GRBs 050603, 050714A
and 050827 with a greater delay
than that for GRB~050911, yet X-ray afterglows were detected in these cases.

{\it Swift} bursts which were detected by the XRT before quickly fading below
its sensitivity level (i.e., afterglows which would not have been located at
4.6~hr) were all short. (GRB~050421 is the exception to this rule and is
discussed below.) GRB~050509B ($\rm T_{90}\sim 30$~ms; Hurkett et
al. 2005; Barthelmy et al. 2005c; Gehrels et al. 2005) and
GRB~050813 ($\rm T_{90} \sim 0.6~\rm s$; Sato et al. 2005; Fox et al. 2005c) both
showed no X-ray emission after a few thousand seconds.
For GRB~050906 ($\rm T_{90} \sim 128~\rm ms$; Parsons et
al. 2005) only a possible, extremely faint X-ray counterpart was
identified (Fox et al. 2005a). Likewise, GRB~050925 ($\rm T_{90} \sim 72~\rm ms$;
Holland et al 2005; Beardmore et al. 2005) was not detected after a
prompt slew, although there is a possibility that this BAT trigger was due to a
new soft $\gamma$-ray repeater.

Comparison of our stringent early X-ray upper limit for GRB~050911 of
$5.2\times 10^{-14}$ $~\rm erg~cm^{-2}~s^{-1}$ with the light-curves of other {\it Swift} bursts
(e.g., fig. 2 of Nousek et al. 2005) shows that the afterglow must have
been an order of magnitude or more fainter than all of the other long bursts at 4.6~hr, with the possible exception of
GRB~050421. The light-curve of that burst (Godet et al. 2005) appears to be
only the tail-end of the prompt emission, with no evidence for an
afterglow (probably due to the burst occuring in a low density
environment: a `naked' GRB). 

The ratio between the 4.6~hr-XRT and $\rm T_{90}$-BAT fluxes 
 was measured for a large number of
bursts and the lowest value found to be $\sim 4 \times 10^{-6}$. Considering the GRB~050911 BAT
flux ($1.86\times 10^{-8}~\rm erg~cm^{-2}~s^{-1}$), this lowest observed ratio
would predict an X-ray flux of $\sim 7\times 10^{-14}~\rm erg~cm^{-2}~s^{-1}$ at 4.6~hr, higher than the upper limit
determined. Thus, GRB~050911 was particularly X-ray faint for the measured
$\gamma$-ray flux.

Our Gemini limit on the optical flux of GRB~050911 is amongst the deepest obtained for a long duration
{\it Swift} burst; only the afterglow of GRB~050412 was apparently
fainter than this, with a limit of $Rc$ = 24.9 only two~hr after
the burst (Kosugi et al. 2005). However
optical afterglows for the recent short bursts GRBs 050509B and
050813 were undetected to deep limits via Keck, Gemini and VLT
observations  (e.g. Bloom et al. 2005; Hjorth et al. 2005b; Bloom
2005; Berger \& Gladders 2005). Deep observations of the handful
of short bursts which were localised prior to {\it Swift} also
failed to locate any afterglow candidates (e.g. Hurley et al. 2003;
Klotz et al. 2003).

\subsection{Long or short?}

From the discussion above, there are at least two possibilities for GRB~050911: either the light-curve is
similar to GRB~050421, showing no sign of emission due to a forward
shock, or it is akin to those in the short burst class. 
The calculated $\rm T_{90}$ for GRB~050911 falls within the long burst
category, although the initial two peaks in the BAT light-curve are both short
($\sim0.5~\rm s$ each). 

The BAT spectrum is not particularly hard. Short
bursts tend to be spectroscopically harder than the long bursts; the
spectral distinction not strong, though, and
the photon index measured for GRB~050911 is within the range observed for both populations
(e.g. Ghirlanda, Ghisellini \& Celotti 2004).
However, any X-ray emission was unusually faint,
more in keeping with measurements of short bursts (Fox et al. 2005b), which tend to fade below the XRT sensitivity within a few thousand
seconds. 

Both the short bursts GRBs~050509B and 050724 (Gehrels et al. 2005; Barthelmy et
al. 2005) were found to be associated with non-star-forming elliptical
galaxies (thus a compact binary merger origin is more
likely than a hypernova). In the case of GRB~050911, the cluster EDCC~493
(Lumsden et al. 1992; $\rm z = 0.16$)
is in the line of sight (Berger 2005) although without a more
refined position than that from the BAT, it cannot be claimed with any certainty that
the burst is associated with any of the galaxies within the cluster. EDCC~493
is not a rich cluster and a rough estimate suggests that, with a radius of
$\sim0.2\deg$, the chance of such a cluster intersecting the line of sight
is a few percent, so the alignment may just be due to chance. No information
about the star-formation rate in this cluster is known.

If GRB~050911 is, as $\rm T_{90}$ suggests, a long burst, then it is unusually
X-ray faint. It could be due to a collapsar, with the
progenitor initially in a binary system. Assuming the progenitor to be the
secondary star in the system, it could receive a kick into a
low-density environment when the primary star undergoes a supernova explosion. If a GRB forms with very little surrounding
inter-stellar material, only a weak forward shock will occur and therefore little afterglow emission
will result.

If, instead, GRB~050911 is interpreted as a compact object merger, the $\gamma$-ray emission out beyond 10~s has to be
explained; as discussed earlier, the burst is not unique in this respect. Models involving two neutron stars are unlikely to produce the
long and structured emission seen in GRB~050911, because the mass ratio in a NS-NS merger is inevitably close
to unity. The disruption of both neutron stars occurs as a single
event and the light-curve is controlled by the properties of the resulting neutron
torus. However, if the mass ratio is far from unity, the disruption event can be more complex. Davies,
Levan \& King (2005) consider a BH-NS merger with a mass ratio of 10:1.
They find that mass transfer from the neutron star occurs in spurts, with
instantaneous rates exceeding $100~\rm \msun s^{-1}$. Between these spurts
the neutron star remnant travels in a wider, eccentric orbit with an initial size
dependent on how much of the angular momentum of the transferred mass is
returned to the remnant. The orbits decay by gravitational radiation,
causing subsequent bursts. For returned angular momentum fractions $\sim$0.5,
these intervals can be several seconds. If the black hole is formed relatively
early on in this process, delayed accretion events could then cause later
outbursts. When the
neutron star remnant reaches a mass of $\sim 0.2~\rm \msun$ it is completely
disrupted, effectively terminating the accretion. Evidently more work is needed
to model in detail the light-curve predicted by this pattern of mass transfer,
but the luminosities and timescales produced from a BH-NS merger resemble those in GRB~050911.

\vspace{-10pt}

\section{Conclusions}
No X-ray afterglow emission was detected for GRB~050911, starting 4.6~hr
after the burst trigger. Comparison of the upper limit of the emission with other
{\it Swift}-detected bursts (Nousek et al. 2005) demonstrates that any
afterglow was at least an order of magnitude fainter than for any other long burst, with the
possible exception of GRB~050421. The behaviour could be due to either
a BH-NS merger, or by a collapsar occuring in a region of low density, thus
forming a `naked' GRB. Whatever mechanism produced GRB~050911, it was an unusual, X-ray dark, burst.

\vspace{-10pt}

\section{Acknowledgments}

The authors acknowledge support for this work at the University of
Leicester by PPARC, at PSU
by NASA and in Italy by ASI. AJL and NRT ackowledge receipt of
PPARC fellowships. We also thank the referee for useful comments and J. Norris
for the spectral lag calculation.


\end{document}